\documentstyle[12pt]{article} 

\begin{document}

\begin{flushright}
TIT/HEP-249/NP \\
April, 1994 \\
\end{flushright}

\begin{center}
{\huge H dibaryon in the QCD sum rule \\}
\vspace*{1.5cm}
{\large Nobuaki Kodama\footnote{e-mail address: kodama@phys.titech.ac.jp}, Makoto Oka\footnote{e-mail address: oka@phys.titech.ac.jp} and Tetsuo Hatsuda* \\
         Department of Physics, Tokyo Institute of Technology\\
         Meguro, Tokyo 152 Japan\\
         \({}^{\ast}\)Institute of Physics, University of Tsukuba\\
        Tsukuba, Ibaraki 305 Japan}
\end{center}
\vspace*{2cm}

\begin{abstract}
The QCD sum rule is applied to the H dibaryon and is compared to the flavor non-singlet di-nucleon. We find that the H dibaryon is almost degenerate to the di-nucleon in the $SU(3)_{flavor}$ limit and therefore is not deeply bound as far as the threshold parameter is adjusted not to have a di-nucleon bound state. After introducing the $SU(3)_{f}$ breaking effects, the H dibaryon is found to be bound by $40 MeV$ below the $\Lambda \Lambda$ threshold.
\end{abstract}
\newpage
\section{Introduction}
\hspace*{0.5cm}
The QCD sum rule is a powerful tool in studying the hadron mass spectrum and other properties directly from QCD\cite{SVZ}. The sum rule is based on the analytical continuation of a current correlation function between the nonperturbative physical region and the perturbative high $Q^{2}$ region. On the theoretical side, the current correlation function is calculated perturbatively for a large Euclidean momentum carried by the current and then the result is analytically continued to the physical spectrum region. On the phenomenological side, the physical spectral density is parameterized with a discrete pole term and a continuum part, whose parameters (the position and the strength of the pole, the position of the threshold, e.t.c.) are determined so as to coincide with the theoretical one. The process of the analytical continuation may often be subtle, because the pole parameters are usually not so sensitive to the short-distance behavior of the correlation function. The Bore!
l sum technique is popularly empl

In this paper we apply the QCD sum rule to the H dibaryon, which is a six-quark bound state predicted by Jaffe in the MIT bag model\cite{jaffe}. Many attempts to calculating the mass of the H dibaryon have been done using various effective theories\cite{Inst}. Yet the result is not conclusive. We apply the QCD sum rule to this problem and calculate the mass of H directly from QCD.

In sect.2, we study how to choose the H dibaryon current and calculate the H dibaryon sum rule in the $SU(3)_{f}$ limit. In sect.3, we calculate the di-nucleon sum rule with the same approximation. In sect.4, we compare the H dibaryon sum rule with the di-nucleon one. In sect.5, we calculate the H dibaryon mass with the $SU(3)_{f}$  breaking effect. In sect.6, we discuss and conclude this work.

\section{ H dibaryon Sum Rule }
\hspace*{0.5cm}
In this section, we consider the sum rule for the H dibaryon. The H dibaryon is a six-quark bound state with $I=0, J=0$ and strangeness $-2$ and is supposed to belong to the $SU(3)_{f}$ singlet representation. Jaffe predicted a bound state with $ 80MeV$ binding energy in the MIT bag model\cite{jaffe}. The binding energy comes mainly from the magnetic gluon exchange between the valence quarks, which is most attractive for the flavor singlet state of six quarks. We construct the interpolating current for the H dibaryon as a product of two baryonic currents. First we consider the $SU(3)_{f}$ symmetric limit with $m_{u}=m_{d}=m_{s}=0$. We employ as the baryon current a linear combination of two currents, one with the nonrelativistic limit and the other without:
\begin{equation}
J_{B}^{p}(x)
= \epsilon_{abc} \epsilon_{ijk} \lambda_{kl}^{p} \cdot
[ (q^{Ta}_{i}(x)Cq^{b}_{j}(x)) \gamma_{5}
q^{c}_{l}(x)+ t (q^{Ta}_{i}(x)C\gamma_{5}q^{b}_{j}(x))q^{c}_{l}(x)]
\label{eqn : bacur}
\end{equation}
where $q$ is the quark field, $i,j,k,l$ are the flavor indices, $a,b,c$ are the color indices and $C$ stands for the charge conjugation operator. $\lambda^{p}$ is the generator of $SU(3)$. 
According to ref.\cite{espriu}, the optimum combination for the baryon sum rule is $t=-1.1$, while the current used by Ioffe corresponds to $t=-1$\cite{Ioffe}.
Using the current (\ref{eqn : bacur}) we construct the H dibaryon current,
\begin{eqnarray}
J_{H}(x)&=&J_{B}^{p} C \gamma_{5} J_{B}^{p} \nonumber \\
 &=& \epsilon_{abc} \epsilon_{a'b'c'} (2 \epsilon_{ijk'}
 \epsilon_{i'j'k} - \frac{2}{3} \epsilon_{ijk}
 \epsilon_{i'j'k'}) \cdot \nonumber \\
&&[(q^{Ta}_{i}(x)Cq^{b}_{j}(x))q^{Tc}_{k}(x)
\gamma_{5} + t (q^{Ta}_{i}(x)C\gamma_{5}q^{b}_{j}(x))
q^{Tc}_{k}(x)]C\gamma_{5}\cdot \nonumber \\
&&[\gamma_{5}q^{c'}_{k'}(x) 
(q^{Ta'}_{i'}(x)Cq^{b'}_{j'}(x)) + t q^{c'}_{k'}(x)
(q^{Ta'}_{i'}(x)C\gamma_{5}q^{b'}_{j'}(x))].
\label{eqn : hcurr}
\end{eqnarray}
Here the flavor index $p$ is summed up so that $J_{H}$ belongs to the flavor singlet representation. The flavor combination factor comes from the formula,
$$\mathop{\Sigma}_{p} \lambda^{p}_{ij} \lambda^{p}_{kl} = 
2\delta_{il}\delta_{jk}-\frac{2}{3} \delta_{ij}\delta_{lk}.$$
The current (\ref{eqn : hcurr}) creates a color-singlet and flavor-singlet 6-quark (uuddss) system and is supposed to the H dibaryon state. 

The operator product expansion (OPE) is applied to the current correlation function,
\begin{equation}
\Pi_{H}(q^2) \equiv -i \int d^{4}x e^{iqx} \langle 0| T\ J_{H}(x)\ 
J_{H}^{\dagger}(0) |0 \rangle 
\end{equation}
for large $Q^2=-q^2$. We make the following assumptions.

\begin{enumerate}
\item {We neglect the gluon condensate, taking only the quark condensate, $\langle\overline{q}^{2}q^{2}\rangle$, and $\langle\overline{q}^{4}q^{4}\rangle$. Although $\langle\alpha_{s} G^{\mu\nu} G_{\mu\nu}\rangle$ has a lower dimension than $\langle\overline{q}q\rangle^{2}$, it is of higher order in $\alpha_{s}$ and has an extra suppression factor $1/(2\pi)^4$.} 
\item {The vacuum saturation is assumed in $\langle\overline{q}^{2}q^{2}\rangle$ and $\langle\overline{q}^{4}q^{4}\rangle$, i.e., $\langle\overline{q}^{2}q^{2}\rangle \sim \langle\overline{q}q\rangle^2$, $\langle\overline{q}^{4}q^{4}\rangle \sim \langle\overline{q}q\rangle^4$. This assumption can be justified in the large $N_{c}$ limit, though in reality ($N_{c}=3$) it has been suggested that this may not be a good approximation\cite{vacuum}.}
\end{enumerate}

The correlation function in the $SU(3)_{f}$ limit is given by
\begin{eqnarray}
\Pi_{H}(q^2)&=& \frac{h1(t)}{2^{14}\pi^{10}\Gamma(9)\Gamma(8)}
(-q^{2})^{7}ln(-q^{2})
+\frac{h2(t)}{2^{8}\pi^{6}\Gamma(6)\Gamma(5)}
(-q^{2})^{4}ln(-q^{2}) \cdot \frac{\langle\overline{q}q\rangle^{2}}{(4N_{c})^{2}} \nonumber \\
&&+\frac{h3(t)}{2^{2}\pi^{2}
\Gamma(3)\Gamma(2)}(-q^{2})ln(-q^{2})
\cdot \frac{\langle\overline{q}q\rangle^{4}}{(4N_{c})^{4}}
\label{eqn : hcorr}
\end{eqnarray}
where,
\begin{eqnarray}
h1(t)&=&60928+186368 t +218112 t^2 + 88064 t^3 +110080 t^4 \nonumber \\
h2(t)&=&243712+745472 t +872448 t^2 + 352256 t^3 -2213888 t^4 \nonumber \\
h3(t)&=&-974848-2981888 t -3489792 t^2 -1409024 t^3 +8855552 t^4 
\end{eqnarray}
 For the phenomenological side we assume the spectral density with a pole (a narrow and the lightest resonance in this channel) plus a continuum. The continuum part is assumed to coincide with the imaginary part of the perturbative $\Pi_{H}(q^2)$ with a step-function cut off at the threshold $s_{0}$:
\begin{eqnarray}
\lefteqn{\frac{1}{\pi}\mbox{Im}\Pi_{H}(s)} \nonumber \\
&=& \lambda^{2}_{H}\delta(m_{H}^{2}-s) +
 [ \frac{h1(t)}{2^{14}\pi^{10}\Gamma(9)\Gamma(8)} s^{7}
 - \frac{h2(t)}{2^{8}\pi^{6}\Gamma(6)\Gamma(5)} s^{4} 
\cdot \frac{\langle\overline{q}q\rangle^{2}}{(4N_{c})^{2}} \nonumber \\
 &&+ \frac{h3(t)}{2^{2}\pi^{2}\Gamma(3)\Gamma(2)} s
\cdot \frac{\langle\overline{q}q\rangle^{4}}{(4N_{c})^{4}} ] \theta(s-s_{0}).
\label{eqn : phen}
\end{eqnarray}
Here, the dispersion relation is given by 
\begin{equation}
\Pi(q^2) = \frac{1}{\pi} \int^{\infty}_{0} ds \frac{\mbox{Im} 
\Pi(s)}{s - q^2 - i \epsilon} + \mbox{"subtractions."} 
\end{equation}
The Borel transformation, 
\begin{equation}
\mbox{L}_{M}=\lim_{\stackrel
     {\stackrel{n\rightarrow \infty}{Q^2\rightarrow\infty}}
     {Q^2/n\rightarrow M^2}} \frac{(-Q^2)^n}{(n-1)!} (\frac{d}{dQ^2})^{n}
\end{equation}
vanishes the subtraction terms and suppresses the continuum contribution.
 
Then the Borel sum rule for the H dibaryon gives
\begin{eqnarray}
\lefteqn{\lambda^{2}_{H}\exp (-m_{H}^{2}/M^{2})} \nonumber \\
&=& \frac{h1(t)}{2^{14}\pi^{10}\Gamma(9)} 
(M^{2})^{8} (1-\Sigma_{7}) - 
\frac{h2(t)}{2^{8}\pi^{6}\Gamma(6)}  
\cdot \frac{\langle\overline{q}q\rangle^{2}}{(4N_{c})^{2}} 
(M^{2})^{5} (1-\Sigma_{4}) \nonumber \\
&&+ \frac{h3(t)}{2^{2}\pi^{2}\Gamma(3)} 
\cdot \frac{\langle\overline{q}q\rangle^{4}}{(4N_{c})^{4}}
(M^{2})^{2}(1-\Sigma_{1})
\label{eqn : hsr}
\end{eqnarray}
where
\[ \Sigma_{ \mbox{i}} =\mathop{\sum}_{k=0}^{i} 
\frac{s_{0}^{k}}{(M^{2})^{k}k!}e^{-s_{0}/M^{2}}. \]
The terms with $\Sigma_{ \mbox{i}}$ represent the continuum part.
The mass of H dibaryon, $m_{H}$, can be extracted by taking the logarithmic derivative of (\ref{eqn : hsr}) with respect to $\frac{1}{M^2}$. We obtain
\begin{eqnarray}
m_{H}^{2} (M^2)
&=& [\frac{h1(t)}{2^{14}\pi^{10}\Gamma(8)} 
(M^{2})^{9} (1-\Sigma_{8}) - 
\frac{h2(t)}{2^{8}\pi^{6}\Gamma(5)}  
\cdot \frac{\langle\overline{q}q\rangle^{2}}{(4N_{c})^{2}} 
(M^{2})^{6} (1-\Sigma_{5}) \nonumber \\
&&+ \frac{h3(t)}{2^{2}\pi^{2}\Gamma(2)} 
\cdot \frac{\langle\overline{q}q\rangle^{4}}{(4N_{c})^{4}}
M^{2}(1-\Sigma_{2}) ]/ \nonumber \\
&& [\frac{h1(t)}{2^{14}\pi^{10}\Gamma(9)} 
(M^{2})^{8} (1-\Sigma_{7}) - 
\frac{h2(t)}{2^{8}\pi^{6}\Gamma(6)}  
\cdot \frac{\langle\overline{q}q\rangle^{2}}{(4N_{c})^{2}} 
(M^{2})^{5} (1-\Sigma_{4}) \nonumber \\
&&+ \frac{h3(t)}{2^{2}\pi^{2}\Gamma(3)} 
\cdot \frac{\langle\overline{q}q\rangle^{4}}{(4N_{c})^{4}}
(M^{2})^{2}(1-\Sigma_{1})]
\label{eqn : hmass}
\end{eqnarray}

 In order to predict $m_{H}$, one has to determine the threshold $s_{0}$. In the standard Borel sum rule, it is fixed by the stability of $m_{H}(M^2)$ in an interval of the Borel mass $M$. This interval is determined by two conditions which will be given in sect.4. In the present case, however, there is no Borel-mass stability achieved. One might be tempted to fix $s_{0}$ using the physical threshold. If we choose twice of the lambda mass (which is degenerate to the nucleon in the $SU(3)_{f}$ limit) for the threshold $s_{0}$, then the H dibaryon seems to be deeply bound. We, however, find that the same parameter makes the flavor non-singlet 6-quark state (di-nucleon) deeply bound as well. The appearance of the unrealistic di-nucleon bound state indicates that the choice of the threshold parameter $s_{0}$ is not appropriate. We therefore will determine $s_{0}$ so that the predicted di-nucleon mass coincides with twice the nucleon mass.

\section{Di-Nucleon Sum Rule}
\hspace*{0.5cm}
In this section we calculate the di-nucleon sum rule under the same approximation with the H dibaryon. The di-nucleon is defined as a six-quark state with strangeness zero and $J=0$, which may couple to pp (or nn) . In order to compare this sum rule with that of the H dibaryon, we assume that the spectrum density has a di-nucleon pole (bound state) and a continuum. 

Similarly to the H dibaryon case, we construct the di-nucleon current as a product of two proton currents. The proton current is given by  
\begin{equation}
J_{p}(x)
= \epsilon_{abc}[ (u^{Ta}(x)Cd^{b}(x)) \gamma_{5}u^{c}(x) + 
t (u^{Ta}(x)C\gamma_{5}d^{b}(x))u^{c}(x)]
\label{eqn : nucur}
\end{equation}
where $u$ is the up-quark field and $d$ is the down-quark field. The di-nucleon current is given by
\begin{equation}
J_{D}(x)=J_{p} C \gamma_{5} J_{p} .
\label{eqn : dicur}
\end{equation}
This current creates a color-singlet but not flavor-singlet 6-quark (uuuudd) system. The correlation function is given by
\begin{eqnarray}
\Pi_{D}(q^2) &=&-i \int d^{4}x e^{iqx} \langle T\ J_{D}(x)\ 
J_{D}^{\dagger}(0)\rangle \nonumber \\
&=& \frac{d1(t)}{2^{14}\pi^{10}\Gamma(9)\Gamma(8)}
(-q^{2})^{7}ln(-q^{2}) 
+\frac{d2(t)}{2^{8}\pi^{6}\Gamma(6)\Gamma(5)}
(-q^{2})^{4}ln(-q^{2}) \cdot \frac{\langle\overline{q}q\rangle^{2}}{(4N_{c})^{2}} 
\nonumber \\
&&+\frac{d3(t)}{2^{2}\pi^{2}
\Gamma(3)\Gamma(2)}(-q^{2})ln(-q^{2})
\cdot \frac{\langle\overline{q}q\rangle^{4}}{(4N_{c})^{4}}
\label{eqn : dcorr}
\end{eqnarray}
where
\begin{eqnarray}
d1(t)&=&135+108 t +234 t^2 + 108 t^3 +135 t^4 \nonumber \\
d2(t)&=&540+432 t +936 t^2 + 432 t^3 -2340 t^4 \nonumber \\
d3(t)&=&-2160-1728 t -3744 t^2 -1728 t^3 +9360 t^4 
\end{eqnarray}
The di-nucleon (D) mass can be obtained in the same procedure as H,
\begin{eqnarray}
m_{D}^{2}
&=& [\frac{d1(t)}{2^{14}\pi^{10}\Gamma(8)} 
(M^{2})^{9} (1-\Sigma_{8}) - \frac{d2(t)}{2^{8}\pi^{6}\Gamma(5)}  
\cdot \frac{\langle\overline{q}q\rangle^{2}}{(4N_{c})^{2}} 
(M^{2})^{6} (1-\Sigma_{5}) \nonumber \\
&&+ \frac{d3(t)}{2^{2}\pi^{2}\Gamma(2)} 
\cdot \frac{\langle\overline{q}q\rangle^{4}}{(4N_{c})^{4}}
M^{2}(1-\Sigma_{2}) ]/ \nonumber \\
&& [\frac{d1(t)}{2^{14}\pi^{10}\Gamma(9)} 
(M^{2})^{8} (1-\Sigma_{7}) - \frac{d2(t)}{2^{8}\pi^{6}\Gamma(6)}  
\cdot \frac{\langle\overline{q}q\rangle^{2}}{(4N_{c})^{2}} 
(M^{2})^{5} (1-\Sigma_{4}) \nonumber \\
&&+ \frac{d3(t)}{2^{2}\pi^{2}\Gamma(3)} 
\cdot \frac{\langle\overline{q}q\rangle^{4}}{(4N_{c})^{4}}
(M^{2})^{2}(1-\Sigma_{1})]
\label{eqn :dmass}
\end{eqnarray}

This equation is very similar to the H dibaryon mass. In the OPE of the current correlation function(\ref{eqn : dcorr}), the second and third terms are identical to those in (\ref{eqn : hcorr}), while the first term has a small difference. This indicates that in the mass calculation only the bare quark loop term ( the first term of the OPE) can make a difference between the H dibaryon and the di-nucleon. Because the first term is much smaller than the other terms, the difference of the H mass and the di-nucleon mass is expected to be small.

\section{Comparing H and Di-Nucleon}
\hspace*{0.5cm}
To proceed with the sum rule, we determine the  interval of the Borel mass $M$ in which the sum rule prediction is reliable. This interval is restricted by the conditions: 1) OPE tends to converge, and 2) the pole contribution is dominant to the continuum one. The lower limit of $M$ is determined  as the value at which the ratio of the first plus the second terms and the third one in OPE is $30\%$. The upper limit is determined as the value where the continuum contribution in the mass prediction is less than $50\%$. These conditions give different intervals of $M$ in the H dibaryon and the di-nucleon sum rules. We, however, take the same upper and lower limits of $M^2$ for the H and the di-nucleon because then difference is small and also our purpose is to compare these two sum rules.

In this paper we adopt $\langle \overline{q} q\rangle=(-0.25GeV)^3$, which leads to the lower limit $M^2 =1.97GeV^2$. When we take $s_{0}=5.7GeV^2$, which results in the di-nucleon mass equal to twice of the nucleon mass, $1.880GeV$, the upper limit is given by $M^2=2.54GeV^2$.

We compare the H dibaryon mass with the di-nucleon one within the above interval of $M$. Although both the sum rules predict no stable plateau of $m$ against $M$, the dibaryon masses depend only weakly on the Borel mass M($<5\%$)(fig.1). It is assumed that the threshold $\sqrt{s_{0}}$ for the H and the di-nucleon sum rules is the same. We compare the masses of the H and the di-nucleon at the central point of the allowed interval ($M^2=2.26GeV^2$) (fig.2), where the coupling strength $\lambda$ is almost stable against the Borel mass(fig.3)($<1\%$). 

From the di-nucleon sum rule, we find that the di-nucleon mass becomes $1.880GeV$, i.e.,  twice of the nucleon mass at $s_{0}=5.694GeV^2$. For this $s_{0}$, the H dibaryon mass is $1.878GeV$, i.e., $m_{di-nucleon}-m_{H}\simeq 2MeV$ .  This remarkable degeneracy is, in fact, not accidental. We find that over a wide range of the Borel mass ($1GeV^2<M^2<\infty$) and the threshold ($4<s_{0}<9 (GeV^2)$), the H dibaryon is almost degenerate or a few $MeV$ below the di-nucleon. The only exception is for around $t=-0.69$ where $\lambda^{2}_{Di-nucleon}$ vanishes and therefore the di-nucleon mass diverges.

This result contrasts sharply from what the quark model suggests. The SU(6) quark model with a gluon exchange interaction yields a distinct flavor-singlet bound state, while flavor non-singlet 6-quark states lie above the two-baryon threshold. The binding energy of the flavor-singlet state, typically of the order of $100 MeV$ in the $SU(3)_{f}$ limit, seems insensitive to the choice of the quark model parameters as far as the hyperfine splittings of the hadrons are reproduced. (One possible exception is the instanton induced interaction, which gives an extra three-body repulsion to the H dibaryon\cite{III}.) The present result in the QCD sum rule does not show the splitting of the flavor-singlet and non-singlet 6-quark states.

\section{SU(3) breaking effect}
\hspace*{0.5cm}
Effects of the $SU(3)_{f}$ breaking can be taken into  account by introducing the strange quark mass and the difference of $\langle\overline{s}s\rangle$ from $\langle\overline{u}u\rangle$. We employ the same $SU(3)_{f}$ singlet current (\ref{eqn : hcorr}) for the interpolating current. Then we obtain, for $t=-1.1$,
\begin{eqnarray}
\lefteqn{-i \int d^{4}x e^{iqx} \langle T\ J_{H}(x)\ 
J_{H}^{\dagger}(0)\rangle } \nonumber \\
&=& \frac{1.638 \cdot 10^5}{2^{14}\pi^{10}\Gamma(9)\Gamma(8)}
(-q^{2})^{7}ln(-q^{2}) \nonumber \\
&-&\frac{(-q^{2})^{5}ln(-q^{2})}{2^{10}\pi^{8}\Gamma(7)\Gamma(6)}
\frac{m_{s}}{4N_{c}}(4.214 \cdot 10^5 \langle\overline{u}u\rangle -4.674 \cdot 10^5 \langle\overline{s}s\rangle) \nonumber \\
&-&\frac{(-q^{2})^{4}ln(-q^{2})}{2^{8}\pi^{6}\Gamma(6)\Gamma(5)}
\frac{1}{(4N_{c})^{2}}(1.311 \cdot 10^6\langle\overline{u}u\rangle^{2} +1.686 \cdot 10^6 \langle\overline{u}u\rangle\langle\overline{s}s\rangle+
2.341 \cdot 10^5 \langle\overline{s}s\rangle^{2}) \nonumber \\
&+&\frac{(-q^{2})^{2}ln(-q^{2})}{2^{4}\pi^{4}\Gamma(4)\Gamma(3)}
\frac{m_{s}}{(4N_{c})^{3}}(1.686\cdot 10^6 \langle\overline{u}u\rangle^{3} + 1.311 \cdot 10^6 \langle\overline{u}u\rangle^{2} \langle\overline{s}s\rangle-
8.428\cdot 10^5 \langle\overline{s}s\rangle^{2}\langle\overline{u}u\rangle) \nonumber \\
&+&\frac{(-q^{2})ln(-q^{2})}{2^{2}\pi^{2}\Gamma(3)\Gamma(2)}
\frac{1}{(4N_{c})^{4}}(9.364\cdot 10^5 \langle\overline{u}u\rangle^{4} +6.743\cdot 10^6 \langle\overline{u}u\rangle^{3} \langle\overline{s}s\rangle+
5.024\cdot 10^6 \langle\overline{s}s\rangle^{2}\langle\overline{u}u\rangle^{2}) \nonumber \\
&+&\frac{4}{(-q^{2})}
\frac{m_{s}}{(4N_{c})^{5}}(4.305\cdot 10^6 \langle\overline{s}s\rangle \langle\overline{u}u\rangle^{4} +3.371\cdot 10^6\langle\overline{u}u\rangle^{3} \langle\overline{s}s\rangle^{2}) .
\label{eqn : hcor2}
\end{eqnarray}
Applying the Borel transformation with $\alpha \equiv \frac{\langle\overline{s}s\rangle}{\langle\overline{u}u\rangle} = 0.8$ and $m_{s}=150MeV$, we obtain the mass of H dibaryon:
\begin{eqnarray}
m_{H}^{2}
&=& [\frac{1.638\cdot 10^5}{2^{14}\pi^{10}\Gamma(8)} 
(M^{2})^{9} (1-\Sigma_{8}) - \frac{5.760\cdot 10^4}{2^{10}\pi^{8}\Gamma(6)}  
\cdot \frac{\langle\overline{q}q\rangle}{4N_{c}} 
(M^{2})^{7} (1-\Sigma_{6}) \nonumber \\
&&+ \frac{2.809 \cdot 10^6}{2^{8}\pi^{6}\Gamma(5)}  
\cdot \frac{\langle\overline{q}q\rangle^{2}}{(4N_{c})^{2}} 
(M^{2})^{6} (1-\Sigma_{5}) - \frac{3.293\cdot 10^5}{2^{4}\pi^{4}\Gamma(3)}  
\cdot \frac{\langle\overline{q}q\rangle^{3}}{(4N_{c})^{3}} 
(M^{2})^{4} (1-\Sigma_{3}) \nonumber \\
&&+ \frac{9.687 \cdot10^6}{2^{2}\pi^{2}\Gamma(2)} 
\cdot \frac{\langle\overline{q}q\rangle^{4}}{(4N_{c})^{4}}
(M^{2})^{3}(1-\Sigma_{2}) ]/ \nonumber \\
&&[\frac{1.638 \cdot 10^5}{2^{14}\pi^{10}\Gamma(9)} 
(M^{2})^{8} (1-\Sigma_{7}) - \frac{5.760 \cdot 10^4}{2^{10}\pi^{8}\Gamma(7)}  
\cdot \frac{\langle\overline{q}q\rangle}{4N_{c}} 
(M^{2})^{6} (1-\Sigma_{5}) \nonumber \\
&&+ \frac{2.809 \cdot 10^6}{2^{8}\pi^{6}\Gamma(6)}  
\cdot \frac{\langle\overline{q}q\rangle^{2}}{(4N_{c})^{2}} 
(M^{2})^{5} (1-\Sigma_{4}) - \frac{3.293 \cdot 10^5}{2^{4}\pi^{4}\Gamma(4)}  
\cdot \frac{\langle\overline{q}q\rangle^{3}}{(4N_{c})^{3}} 
(M^{2})^{3} (1-\Sigma_{2}) \nonumber \\
&&+ \frac{9.687 \cdot 10^6}{2^{2}\pi^{2}\Gamma(3)} 
\cdot \frac{\langle\overline{q}q\rangle^{4}}{(4N_{c})^{4}}
(M^{2})^{2}(1-\Sigma_{1}) + 4 \cdot 1.929 \cdot 10^5
\cdot \frac{\langle\overline{q}q\rangle^{5}}{(4N_{c})^{5}} ]
\end{eqnarray}

In evaluating the mass of H, we assume that the threshold $\sqrt{s_{0}}$ is twice of the strange quark mass ($=150MeV$) plus $\sqrt{s_{0}^{SU(3)_{f}limit}} (=\sqrt{5.694} GeV)$. The interval of the Borel mass is determined by the same conditions as in the $SU(3)_{f}$ limit, and is given by $M^2=2-3.4GeV^2$. The H mass is then estimated to be $m_{H} \simeq 2.19GeV$. Because the $\Lambda \Lambda$ threshold is $2.23 GeV$, about $40 MeV$ higher than this prediction, we predict a bound state.

\section{Discussion and Conclusion}
\hspace*{0.5cm}
Experimental study of the strangeness $-2$ system has been
actively carried out these few years.  Yet, we have no evidence of the H dibaryon. In fact, the binding energies of the reported double hypernuclei
yield the upper limit of the binding energy of H, $E_{B} \leq 28$
MeV\cite{Hsearch}. 

The present sum rule result predicts a slightly deeper bound state. The number, however, should not be taken too seriously, as the result is extremely sensitive to the choice of the continuum threshold $s_{0}$. Some more ambiguities may come from the corrections due to the neglected terms of the operator product expansion and from the approximations in this calculation, such as the vacuum saturation in the $\langle \overline{q}q \rangle^4$ matrix element. Further studies would be necessary to make a conclusive result from the QCD sum rule. Nevertheless, the present comparison between H and di-nucleon suggests that the binding of H is not as strong as expected in quark model. Especially, the results in the $SU(3)_{f}$ limit do not show any extra attraction for the flavor-singlet state, although the $SU(3)_{f}$ breaking effect seems to enhance the binding energy. These features are clearly different qualitatively from the quark model prediction.

Recent study of the H dibaryon mass in the quark model shows
that the instanton induced interactions, which represent the U(1) anomaly in the meson spectrum, make the H dibaryon less bound\cite{III}.  In the sum rule the same effect could be introduced as the direct instanton contribution on the theoretical side. A resent study suggests a significant role of the direct instanton contribution in the nucleon sum rule \cite{Forkel}. Its contribution to the H dibaryon sum rule is under investigation.

\newpage
{\large Figure caption\\}

\begin{tabular}{cp{12cm}}
Fig.1 & Masses of the H and the di-nucleon v.s. the Borel mass $M^2$. The solid line corresponds to H dibaryon and the dashed line corresponds to di-nucleon.\\
Fig.2 & Masses of the H and the di-nucleon v.s. t.\\
Fig.3-(a) & The coupling $\lambda$ for the H dibaryon.\\
Fig.3-(b) & The coupling $\lambda$ for the di-nucleon.\\
\end{tabular}
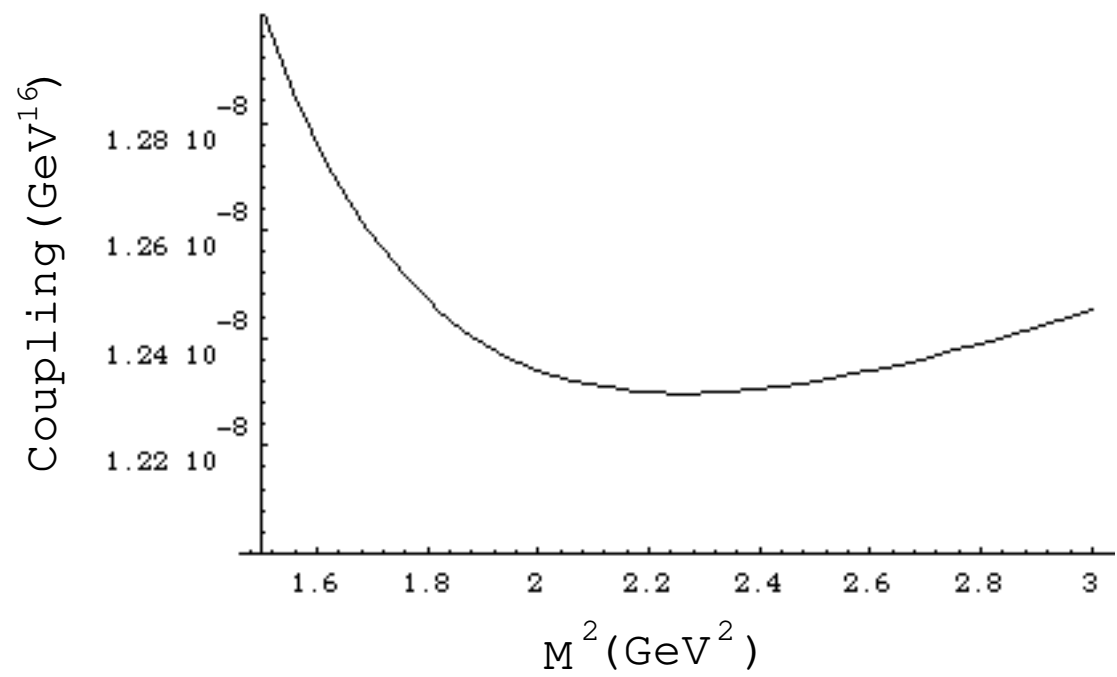
\end{document}